\documentclass[12pt,a4paper]{article}
\usepackage{amsmath}
\usepackage{amssymb}
\usepackage{times}
\usepackage[margin=2.5cm]{geometry}
\usepackage[round]{natbib}
\usepackage{varioref}
\usepackage{hyperref}
\newcommand{\half}{\tfrac{1}{2}}
\newcommand{\quarter}{\tfrac{1}{4}}

\title{Numerical Relativity and Asymptotic Flatness}

\author{E.~Deadman\thanks{Email \texttt{e.deadman@damtp.cam.ac.uk}}
  \mbox{  }\&  
J.M.~Stewart\thanks{Email \texttt{j.m.stewart@damtp.cam.ac.uk}}
\\
Department of Applied Mathematics \& Theoretical Physics\\
Centre for Mathematical Sciences\\
Cambridge CB3 0WA}

\date{\today}

\begin{document}
\bibliographystyle{plainnat}
\labelformat{equation}{\textup{(#1)}}

\maketitle

\setlength{\parskip}{1ex plus 0.5ex minus 0.2ex}
\begin{abstract}
  \noindent
It is highly plausible that the region of space-time far from an
isolated gravitating body is, in some sense, asymptotically
Minkowskian. 
However theoretical studies of the full nonlinear theory, initiated by
\citet{BVM62}, \citet{Sachs62} and \citet{NU62}, rely on careful, clever,
\emph{a-priori} choices of chart (and tetrad) and so are not readily
accessible to the numerical relativist, who chooses her/his chart on
the basis of quite different grounds.

This paper seeks to close this gap.
Starting from data available in a typical numerical evolution, we
construct a chart and tetrad which is, asymptotically, sufficiently
close to the theoretical ones, so that the key concepts of Bondi
news function, Bondi mass and its rate of decrease can be
estimated.
In particular these esimates can be expressed in the numerical
relativist's chart as \emph{numerical relativity recipes}. 
\end{abstract}

% %\tableofcontents

\section{Introduction and motivation}
\label{sec:intro}

The two threads which underpin this study of asymptotic flatness are
theoretical and numerical relativity.
We start by reviewing the former.
It is widely believed that the region of space-time far from an
isolated gravitating body is, in some sense, asymptotically  Minkowskian.
Already in 1962 Bondi and his coworkers \citep{BVM62}, \citep{Sachs62}
developed asymptotic expansions for the solution of the full nonlinear
vacuum field equations, leading to a rigorous concept of gravitational
radiation in the far field.
Soon afterwards \citet{NU62} produced an alternative version using the
NP null tetrad formalism \citep{NP62}. 
A key ingredient in this and later work was the careful choice of a
suitable coordinate chart involving a ``retarded time'' coordinate $u$.
Both groups introduced a $(u,r,\theta,\phi)$ chart where 
$\theta$ and $\phi$ were  spherical polar coordinates.
They both developed asymptotic expansions as $r\rightarrow\infty$
holding the other coordinates fixed.
Subsequent work by Penrose \citep{RP63} showed that by a process of
``conformal compactification'', infinity could be adjoined to the
space-time manifold and then treated by standard methods.
Most modern theoretical treatments use the Penrose conformal approach,
and the ``old-fashioned'' chart-based approach has fallen out of
fashion.
However it is closer to what most numerical relativists are
calculating, and for this reason we shall use it here. 

The two groups used different charts and dependent variables.
\citet{BVM62}  chose a $(u,r,\theta,\phi)$ chart where $r$ was an area
coordinate and $\theta$ and $\phi$ were standard spherical polar
coordinates (see below). 
Their primary dependent variables were the metric components.
\citet{NU62} produced an alternative version using a null ``retarded
time''  coordinate $u$.
Then the  null vector $l^{a} = g^{ab}u_{,b}$ is geodesic and their
coordinate $r$ was chosen to be an affine parameter for the integral
curves of $l^{a}$ along which the other three coordinates were fixed.
Their primary dependent variables were the tetrad connection
components and the tetrad components of the Weyl curvature tensor.  
These ten independent Weyl tensor components are usually described by
five complex scalar functions $\Psi_{n}$ where $n=0,1,\ldots ,4$.
(For a  covariant physical interpretation of the Weyl tensor see e.g.,
\citet{Szek65}.) 

Both groups  were considering the limit $r\rightarrow\infty$ with the
other coordinates fixed.  
In the Penrose geometrical picture \citep{RP63} this region is called
\emph{future null infinity}.
Of course both groups could have considered an ``advanced time''
coordinate $v$ where the corresponding limit is \emph{past null
  infinity}.
The $\Psi_{n}$ in that case have similar properties and interpretation
to $\Psi_{4-n}$ near future null infinity.

If one wants to consider an isolated system with no extrinsic incoming
radiation, then, as explained below, the  natural place to impose this
is past null infinity. 
However both groups looked for a condition to be imposed near future
null infinity. 
\citet{BVM62} introduced an ``outgoing radiation condition'' which
required the vanishing of certain terms in the asymptotic expansion of
two of the metric components. 
This condition will be stated more precisely in section \ref{sec:rett}.
\citet{NU62} made a ``peeling assumption'': near future null infinity 
$\Psi_{0} = O(r^{-5})$, and with this assumption they were able to
demonstrate a so-called ``peeling theorem'':  $\Psi_{n}= O(r^{n-5})$.
Then $\Psi_{4} = O(r^{-1})$ is interpreted as the leading term in the
outgoing radiation.
In the  \citep{BVM62} picture the equivalent r\^ ole is
taken by the ``Bondi news function'' built from first derivatives of
metric components.
For a comparison of the conditions in the two schemes, showing that
the outgoing radiation condition implies the peeling assumption
see e.g., \citet{Kroon98}, and section \ref{sec:curvature}.

By reversing the direction of time, swapping advanced time for
retarded time, one could carry out an almost identical study near past
null infinity.
There, assuming the analogous peeling condition, $\Psi_{0} =
O(r^{-1})$ is to be interpreted as the leading term in the extrinsic
incoming radiation, and a natural ``no incoming radiation condition''
near past null infinity would be $\Psi_{4}=O(r^{-5})$.  

It is important to realise that the ``outgoing radiation condition''
or ``peeling assumption'' does not preclude the presence of incoming
radiation near future null infinity.
Even within linearized theory the ``peeling theorem'' allows modest
amounts of incoming radiation (Deadman \& Stewart, paper in preparation).

We turn now to numerical relativity where researchers  have expended
considerable effort on the (numerical) evolution of asymptotically
flat space-times. 
A minority of researchers have adopted the Penrose conformal approach,
but most have chosen to evolve the space-time as far out (both in
space and time) as is feasible, using the traditional approach.
Then some matching process is required to interpret their numerical
data in the Bondi or NP pictures.
This, the goal of this paper,  turns out to be far from trivial. 
The choice of a coordinate chart is an intrinsic part of the numerical
evolution and the final data is available only in this chosen chart.
Each numerical relativity group has its own favoured chart or charts
and they usually  bear little resemblance to the Bondi or NP ones.
Furthermore this data does not contain complete information because 
the inevitable occurrence of numerical errors will corrupt the values
of higher derivatives---from it one can construct reliably only a few
leading terms in asymptotic expansions\footnote{Consider an asymptotic
  expansion $$f(r)=f_{0} + f_{1}r^{-1} + f_{2}r^{-2}+\ldots$$ as
  $r\rightarrow\infty$.  If we interpret this as the first few terms in
  the Taylor series for $f(q)$ about $q=0$ where $q=r^{-1}$, then the
  $f_{n}$ are, up to numerical factors, the $q$-derivatives of $f$
  evaluated at $q=0$.}.

The usual approach adopted by numerical relativists is to argue that
far from the isolated source the gravitational field is weak, and so
linearized theory can be used to match the numerical and the Bondi
or NU pictures.
\citet{BVM62} argued strongly against such an approximation pointing
out the fundamental nonlinearity of general relativity.
Even if plausible arguments in its favour could be found, linearization
carries its own difficulties. 
The first is that in a non-compactified space-time the matching
process is a global one.   
Further given a space-time, the choice of a simpler second space-time
of which the first can be considered a linearized perturbation, is not
unambiguous. 
Even if such a choice could be justified the transformation between the
charts in the two space-times would not, in general, be smooth.

As a concrete example illustrating these points consider the
well-known Schwarzschild metric in the standard $(t,r,\theta,\phi)$
chart 
\begin{equation}
  \label{eq:schwarz1}
  g^S_{ab} = \text{diag}(F, -F^{-1}, -r^2, -r^2\sin^2\theta),
\end{equation}
where $F = 1 -2M/r$.
In the region where $r\gg M$ this might appear to be a small
perturbation of Minkowski space-time with metric
\begin{equation}
  \label{eq:mink1}
  g^M_{ab} = \text{diag}(1, -1, -r^2, -r^2\sin^2\theta),
\end{equation}
but this is deceptive.
Consider the scalar wave equation $g^{ab}\Psi_{;ab}=0$ on the two
space-times.
We would measure outgoing radiation at \emph{future null infinity} by
taking the limit $r\rightarrow\infty$ holding $u$ constant, where $u$
is a retarded time coordinate.
Two standard choices for $u$ are
\begin{equation}
  \label{eq:ret1}
  u^M = t - r, \qquad u^S = t - r^*,
\end{equation}
where 
\begin{equation}
  \label{eq:rstar1}
  r^* = \int F^{-1}\,{\rm d}r = r + 2M\log|r/2M - 1| + const.
\end{equation}
Thus
\begin{equation}
  \label{eq:ret2}
  u^M = u^S + 2M\log|r/2M - 1| + const.
\end{equation}
The Schwarzschild null infinity is given by $r\rightarrow\infty$
holding $u^S$ constant, which implies $u^M\rightarrow \infty$, known
as \emph{future timelike infinity} for the Minkowski space-time.
Equivalently the Minkowski null infinity involves taking the limit
with $u^M$ constant which corresponds to $r\rightarrow\infty$ with 
$u^S\rightarrow -\infty$, known as \emph{spacelike infinity}
for the Schwarzschild spacetime.
Thus the limits in the two charts are different.
This happens because of the global nature of the limiting process.

In order to achieve comparable  limiting processes we need to redefine
the two charts.
Here both space-times are static and so it is simplest to retain the
$t$-coordinate.
Suppose we invert (for $r>M$) the relation \ref{eq:rstar1},
$r^*=r^*(r)$ giving $r=r(r^*)$ and introduce a new chart
$(t,r^*,\theta,\phi)$. 
Then the Schwarzschild line element \ref{eq:schwarz1} becomes
\begin{equation}
  \label{eq:schwarz2}
  g^S_{ab} = \text{diag}(F, -F, -r^2, -r^2\sin^2\theta).
\end{equation}
Using the same chart the Minkowski line element is 
\begin{equation}
  \label{eq:mink2}
  g^M_{ab} = \text{diag}(1, -1, -r^2, -r^2\sin^2\theta).
\end{equation}
Now the two metrics \ref{eq:schwarz2} and \ref{eq:mink2} are not
only small perturbations of each other (for large $r^*$), but they
share the same causal structure, $u=t-r^*$ in both cases.
(There are of course many other ways of doing this, e.g., retain the
$r$'s and change the $t$'s, which is the approach to be adopted in
this paper.)
Note also the appearance of logarithms, which  means that the
transformations are not smooth.

The purpose of this paper is to examine in more detail these issues
from the point of view of the numerical relativist.  
In section \ref{sec:numerical} we state what information we believe is
available in a typical numerical evolution, and we assume that this
information is expressed in terms of a given chart  
$X^{a}=(T,R,\Theta,\Phi)$ which is asymptotically Minkowskian.
Section \ref{sec:rett} addresses the construction of an approximate
Bondi-like chart $x^{a}=(u,r,\theta,\phi)$ using this information.
This circumvents the problem referred to above.
We write down here the explicit form of the Bondi et al.~outgoing
radiation condition.
We introduce a Newman-Penrose (NP) tetrad \citep{NP62} adapted to the
problem by \citet{NU62} in section \ref{sec:tetrad}.
At leading order this is the usual NP tetrad for Minkowski space-time.
At each order, $r^{-1}$, $r^{-2}, \ldots ,$ there are 16 real
coefficients describing the tetrad.
However from section \ref{sec:rett} we know that only 10 coefficients
are needed to describe the metric.
There are six coefficients which describe an infinitesimal
Lorentz transformation at each order, and, for the moment, we do not
make a particular choice for them.

In section \ref{sec:curvature} we first obtain the asymptotic solution of
the full nonlinear vacuum Einstein equations.
As we do so we fine-tune our chart and NP tetrad to make them closer
to those of Bondi et al.~and Newman-Unti.
Once we have set the Ricci curvature, to the best of our abilities, to
zero we turn to the Weyl curvature described by the Weyl scalars
$\Psi_{n}$ referred to earlier. 
We find that $\Psi_{n}=O(r^{n-5})$ for $n=4,3,2,1$, but
$\Psi_{0}=O(r^{-4})$, which would appear to violate the NU peeling
assumption. 
However using the information gleaned from solving the vacuum field
equations and the fine-tuning of the chart and tetrad, we can show that
the  Bondi outgoing radiation condition implies the NU peeling
condition so that the peeling theorem then holds.

The bad news is that the leading terms in the Weyl scalars
$\Psi_{0}=O(r^{-5})$ and $\Psi_{1}=O(r^{-4})$ cannot be estimated
using the information we judge to be available from the information
extracted in section \ref{sec:numerical}.
Although these scalars can be computed in linearized theory, that
theory would appear to be an unreliable guide here near future null
infinity.  

The good news is that we can compute the leading terms in
$\Psi_{4}=O(r^{-1})$, equivalent to the ``Bondi news function'' (and
we can compute this scalar accurately within linearized theory).
The same holds for $\Psi_{3}=O(r^{2})$, which involves nonlinear
terms, but these can be removed by the fine-tuning process.
We can also compute $\Psi_{2}= O(r^{-3})$ which involves nonlinear
terms in an essential way.
This means that we can offer reliable estimates of the ``Bondi mass''
$M_{B}(u)$ of the isolated system\footnote{The ``Bondi mass'' is of
  course the timelike component of a 4-vector and so frame dependent.
  But a numerical relativity evolution singles out a well-defined frame,
  and that is the one in which the mass is computed.}, and its rate of
decrease $dM_{B}/du\leqslant0$, presumably due to the radiation of
energy, both manifestly inaccessible to linearized theory.

The final section \ref{sec:numrec} translates these results back into
the $X^{a}$ chart of section \ref{sec:numerical} used by a typical
numerical relativist. 
From her/his standpoint there is no need to go through the elaborate
construction of a theoretical chart and NP tetrad carried out in the
intermediate sections.
We offer ``numerical relativity recipes'' so that they can compute the
key quantities referred to in the previous paragraph in their own
preferred chart.

The key ideas in this paper are at least forty years old, and one
might ask why were these results were not given before?
The nonlinear calculations of \citet{BVM62}, \citet{Sachs62} and
\citet{NU62} were made possible by careful, clever, a-priori choices of
chart and tetrad.
We have to start from more or less arbitrary choices and so the
resulting expressions are  horrendously complicated. 
In order to handle them accurately we have utilized a
computer algebra system.
We used \emph{Reduce}, and our \emph{Reduce 3.8} scripts can be
obtained by an email request to the authors.
Our choice reflected our experience and knowledge of one particular
computer algebra system, but we used no features not available in
some other systems.

We are grateful to Oliver Rinne for very useful discussions.
One of us (ED) is grateful for financial support from the UK
Engineering and Physical Sciences Research Council.

%%% Local Variables: 
%%% mode: latex
%%% TeX-master: "master"
%%% End: 

\section{The numerical data}
\label{sec:numerical}

Most numerical relativists would choose a quasi-spherical polar chart
$X^{a} = (T, R, \Theta, \Phi)$ for the numerical evolution of the
space-time surrounding an isolated gravitational source.
We could also define an associated quasi-cartesian chart $Y^{a} =
(T,X,Y,Z)$ where
\begin{equation*}
  X = R\sin\Theta\cos\Phi, \quad
  Y = R\sin\Theta\sin\Phi, \quad
  Z = R\cos\Theta .
\end{equation*}

We shall be interested in the limit $R\rightarrow\infty$.
As stated this limit is meaningless unless we specify the behaviour of
the other three coordinates under the limiting process, and we shall
rectify this omission shortly.
It proves very convenient to introduce the notation
\begin{equation}
  \label{eq:onotn}
  O_{n} = O(R^{-n})  \text{  as  } R\rightarrow\infty.
\end{equation}
Our fundamental assumption is that the space-time outside an isolated
source is asymptotically Minkowskian, expressed by the idea that, as
seen in the $Y^{a}$ chart,
\begin{equation}
  \label{eq:mink1}
  g_{ab} = \eta_{ab} + g^{(1)}_{ab}R^{-1} + g^{(2)}_{ab}R^{-2} + O_{3},
\end{equation}
where $\eta_{ab}=\text{diag}(1,-1,-1,-1)$ and the $g^{(n)}_{ab}$ are
supposed to remain constant during the limiting process.
Transforming from the Minkowskian chart to the spherical polar one we
find that the metric components in the $X^{a}$ chart look like
\begin{equation}
  \label{eq:gtrdd}
  \begin{split}
    g_{00} &= 1 + h_{00}R^{-1} + k_{00}R^{-2} + O_{3},\\
    g_{01} &=  h_{01}R^{-1} + k_{01 }R^{-2} + O_{3},\\
    g_{02} &= h_{02} + k_{02}R^{-1}  + O_{2},\\
    g_{03} &= h_{03} + k_{03}R^{-1}  + O_{2},\\
    g_{11} &= -1 +  h_{11}R^{-1} + k_{11}R^{-2}  + O_{3},\\
    g_{12} &= h_{12} +  k_{12}R^{-1} + O_{2},\\
    g_{13} &= h_{13} + k_{13}R^{-1}  + O_{2},\\
    g_{22} &= -R^{2} +h_{22} R  + k_{22} + O_{1},\\
    g_{23} &= h_{23} R  + k_{23} + O_{1},\\
    g_{33} &= -R^{2}\sin^{2}\Theta +  h_{33}R + k_{33} + O_{1}.\\
  \end{split}
\end{equation}
Here the functions $\{h_{ab}\}$ and $\{k_{ab}\}$ are required to remain
constant during the limiting process.

We shall also need the asymptotic form of the inverse metric $g^{ab}$
which is readily obtained from the relation 
$g^{ac}g_{cb} = \delta^{a}{}_{b}$.
We find
\begin{equation}
  \label{eq:gtruu}
  \begin{split}
    g^{00} &= 1 + h^{00}R^{-1}  + k^{00}R^{-2}  + O_{3},\\
    g^{01} &= h^{01}R^{-1}  + k^{01} R^{-2}  + O_{3},\\
    g^{02} &= h^{02}R^{-2}  + k^{02}R^{-3}  + O_{4},\\
    g^{03} &= h^{03}R^{-2}  + k^{03}R^{-3}  + O_{4},\\
    g^{11} &= -1 + h^{11}R^{-1}  + k^{11}R^{-2}  + O_{3},\\
    g^{12} &= h^{12}R^{-2}  + k^{12}R^{-3}  + O_{4},\\
    g^{13} &= h^{13}R^{-2}  + k^{13}R^{-3}  + O_{4},\\
    g^{22} &= -R^{-2} + h^{22}R^{-3}  + k^{22}R^{-4}  + O_{5},\\
    g^{23} &= h^{23}R^{-3}  + k^{23}R^{-4}  + O_{5},\\
    g^{33} &= -R^{-2}\csc^{2}\Theta + h^{33}R^{-3}  + k^{33}R^{-4}  + O_{5}.\\
  \end{split}
\end{equation}
Explicit formulae for $h^{ab}$ and $k^{ab}$ are given by equations
\ref{eq:huu} and \ref{eq:kuu} in appendix \ref{sec:appa}.
At this level of approximation
\begin{equation*}
  g^{ac}g_{cb} = \delta^{a}{}_{b} + O_{3}.
\end{equation*}

A numerical evolution in which the dependent variables include both
$g_{ab}$ and $g_{ab,c}\,$, usually called a ``first order formulation'',
should produce accurate values for  $h_{ab}$ and its first derivatives,
and for $k_{ab}$. 
Otherwise we assume that these variables are available for discrete
ranges of $T$, $\Theta$ and $\Phi$ so that the corresponding
derivatives can be estimated. 
%%% Local Variables: 
%%% mode: latex
%%% TeX-master: "master"
%%% End: 

\section{The Bondi chart}
\label{sec:rett}

Most of the theoretical work which has been done on outgoing
gravitational radiation involves a ``Bondi chart'' 
$(u, r, \theta,\phi)$ in which $u$ is a retarded time
coordinate, see e.g., \citet{BVM62}, \citet{NP62}, \citet{NU62}.

Here we take the viewpoint that the $X^{a} = (T,R,\Theta,\Phi)$ chart
introduced in section \ref{sec:numerical} is the fundamental one in
which, ultimately, all numerical calculations will be performed.
Starting from this chart we need to construct an 
$x^{a} = (u, r,\theta,\phi)$ one which has all the essential features 
of a Bondi chart and we start by studying the function 
$u(T,R,\Theta,\Phi)$.

Because $u$ is a null coordinate it has to satisfy the
\emph{relativistic eikonal equation}
\begin{equation}
  \label{eq:eikonal}
  g^{ab}u_{,a}u_{,b} = 0.
\end{equation}
This is a well-known nonlinear equation with four independent
variables which is exceedingly difficult to solve with any generality.
(Even the restriction of \ref{eq:eikonal} to Minkowski space-time
leads to the surprisingly rich structure of light ray caustics.)
Note  that there is a ``gauge freedom''---if $u$ is a
solution then so is $U(u)$ for any differentiable function $U$.

The standard approach is to specify $u$ on a spacelike hypersurface in   
space-time, and then existence and local uniqueness of $u$ is
guaranteed by standard theorems.
The standard approach is of little utility in this context, for no
obvious choice of data suggests itself, and so we adopt a different
approach.

Consider first the special case of a Minkowski space-time, where
\ref{eq:eikonal} can be rewritten as 
\begin{equation}
  \label{eq:eikmink}
  (u_{,T})^{2} - (u_{,R})^{2} = 
  R^{-2}\left[(u_{,\Theta})^{2} + \csc^{2}\Theta(u_{,\Phi})^{2}\right] = O_{2}.
\end{equation}
Suppose we look for spherically symmetric solutions $u=u(T,R)$.
Setting $\omega=u_{,R}/u_{,T}$ in \ref{eq:eikmink} we find
$\omega^{2}=1$.
Using the gauge freedom mentioned earlier we may impose $u_{,T}=1$ to
find
\begin{equation*}
  du = dT \pm dR,
\end{equation*}
which implies
\begin{equation*}
  u = T \pm R + \text{const}.
\end{equation*}
$T-R$ is called \emph{retarded time} and $T+R$ is called
\emph{advanced time}.

Although  the special case appears trivial it is the key to the general
one.
Within this section only let indices $i,j$ range over $0,1$
and let indices $I,J$ range over $2,3$.
Perusal of the display \ref{eq:gtruu} shows that $g^{ij}$ is $O_{0}$
while both $g^{iJ}$ and $g^{IJ}$ are $O_{2}$.
Thus the eikonal equation takes the form
\begin{equation}
  \label{eq:eikgen1}
  g^{ij}u_{,i}u_{,j} = O_{2},
\end{equation}
which should be compared with \ref{eq:eikmink} above.
As boundary conditions (as $R\rightarrow\infty$) we impose
\begin{equation}
  \label{eq:eikbc}
  u_{,T} = 1 + O_{1}, \qquad u_{,I} = O_{1}.
\end{equation}
This means that the eikonal equation takes the form
\begin{equation}
  \label{eq:eikgen2}
  g^{ij}u_{,i}u_{,j} = O_{3},
\end{equation}
which we can write as a quadratic equation for $\omega=u_{,R}/u_{,T}$,
and choosing the sign appropriate for retarded time we find the
solution
\begin{equation}
  \label{eq:ur1}
  u_{,R} = -(1+2m_{1}/R + 2m_{2}/R^{2})u_{,T} + O_{3},
\end{equation}
where
\begin{equation}
  \label{eq:mass1}
  m_{1} = -\tfrac{1}{4}(h_{00}+2h_{01}+h_{11}),
\end{equation}
and
\begin{equation}
  \label{eq:mass2}
  \begin{split}
    m_{2} =& -\tfrac{1}{16}\big[4k_{00} + 8k_{01} + 4k_{11} +
    (h_{00}-h_{11})^{2} -\\
    &\;\;4(h_{00}+h_{01})^{2} +4(h_{02}+h_{12})^{2} +
    4(h_{03}+h_{13})^{2}\csc^{2}\theta\big].
  \end{split}
\end{equation}
Thus 
\begin{equation}
  \label{eq:du}
  du = u_{,T}\,dT - u_{,T}(1+2m_{1}/R+ 2m_{2}/R^{2})\,dR + O_{3}.
\end{equation}
We leave some freedom in $u$ by setting
\begin{equation}
  \label{eq:ut}
  u_{,T} = 1 + \frac{q_{1}}{R} + \frac{q_{2}}{R^{2}} + O_{3},
\end{equation}
where $q_{1}$ and $q_{2}$ are $R$-independent functions. At the moment
they are arbitrary.  The requirement that the vacuum Einstein
equations hold then determines inter alia $q_1$, see section 6. In our
calculation $q_2$ is not used directly. 

We need next to specify a radial coordinate $r = r(T,R,\Theta,\Phi)$,
and the simplest choice is $r=R$.
This has the great practical advantage that 
$O_{n} = O(R^{-n}) = O(r^{-n})$.
It could be argued that our choice of $r$ is neither the Bondi area
coordinate nor an affine parameter along the outgoing null rays as
favoured by \citet{NU62}.
However since both of those approaches are known to be essentially
equivalent, it would seem that the discussion is not sensitive to the
precise choice of $r$.

Then \ref{eq:du} implies
\begin{equation}
  \label{eq:dT}
  dT = \left((1 - q_{1}/r - (q_{2}-q_{1}{}^{2})/r^{2}\right)\,du + 
    (1+2m_{1}/r + 2m_{2}/r^{2})\,dr + O_{3},
\end{equation}
and so
\begin{alignat}{2}
  \label{eq:jac1}
  \left(\frac{\partial T}{\partial u}\right)_{r} &= 
  1 - \frac{q_{1}}{r} - \frac{q_{2}-q_{1}{}^{2}}{r^{2}} +  O_{3}, 
  \quad \left(\frac{\partial R}{\partial u}\right)_{r} &= 0,\\
  \left(\frac{\partial T}{\partial r}\right)_{u} &= 
  1 + \frac{2m_{1}}{r} + \frac{2m_{2}}{r^{2}} + O_{3},
  \quad\;\; \left(\frac{\partial R}{\partial r}\right)_{u} &= 1.
\end{alignat}

Finally we consider the choice of angular coordinates
$\theta = \theta(T, R, \Theta, \Phi)$ and 
$\phi = \phi(T, R, \Theta, \Phi)$.
We shall require $\theta = \Theta + O_{1}$ and $\phi = \Phi + O_{1}$,
and so the relations, being close to the identity, are invertible.
It is more convenient to posit
\begin{equation}
  \label{eq:ang}
  \Theta = \theta + \frac{y_{2}}{r} + \frac{z_{2}}{r^{2}} + O_{3}, \qquad
  \Phi = \phi + \frac{y_{3}}{r} + \frac{z_{3}}{r^{2}} + O_{3}, 
\end{equation}
where the functions $y_{J}$ and $z_{J}$ do not depend on $r$ but are
otherwise arbitrary.
Equations \ref{eq:ang} are certainly consistent with the boundary
conditions \ref{eq:eikbc}.

We can now specify the limiting process as $r\rightarrow\infty$
holding $u$, $\theta$ and $\phi$ constant.
Thus we are regarding $m_{n}$, $q_{n}$, $y_{J}$, $z_{J}$, $\{h_{ab}\}$ 
and $\{k_{ab}\}$ as functions of $u$, $\theta$ and $\phi$.

Because we know the Jacobian $(\partial X^{a}/\partial x^{b})$ we can
write down the metric components in the $x^{a}= (u,r,\theta,\phi)$
chart
\begin{equation}
  \label{eq:gurdd}
  \begin{split}
    g_{00} &= 1 + a_{00}/r + b_{00}/r^{2} + O_{3},\\
    g_{01} &= 1 + a_{01}/r + b_{01}/r^{2} + O_{3},\\
    g_{02} &= -r{y_{2}}_{,u} + a_{02} + b_{02}/r + O_{2},\\
    g_{03} &= -r{z_{2}}_{,u}\sin^{2}\theta + a_{03} + b_{03}/r + O_{2},\\
    g_{11} &= a_{11}/r + b_{11}/r^{2} + O_{3},\\
    g_{12} &= a_{12} + b_{12}/r + O_{2},\\
    g_{13} &= a_{13} + b_{13}/r + O_{2},\\
    g_{22} &= -r^{2} + a_{22}r + b_{22} + O_{1},\\
    g_{23} &= a_{23}r + b_{23} + O_{1},\\
    g_{33} &= -r^{2}\sin^{2}\theta + a_{33}r + b_{33} + O_{1}.\\
  \end{split}
\end{equation}
Two points should be noted here.
Firstly the leading terms in $g_{02}$ and $g_{03}$, if non-zero,
would violate our notion of an asymptotically Minkowskian space-time,
for they are not present in the standard Minkowski line element.
Thus we need to impose the conditions or ``constraints''
\begin{equation}
  \label{eq:yu}
  {y_{2}}_{,u} = {z_{2}}_{,u} = 0.
\end{equation}
Explicit formulae for $a_{mn}$ in terms of the $h_{mn}$, $q_{1}$,
$m_{1}$ and $y_{2}$, $z_{2}$ (after imposing \ref{eq:yu}) are given as
\ref{eq:add} in appendix \ref{sec:appa}.  
We could also give explicit formulae for the $b_{mn}$ in terms of the
$h_{mn}$, $k_{mn}$, $q_{n}$, $m_{n}$, $y_{n}$ and $z_{n}$but they are
rather lengthy, and are most easily generated using a computer algebra 
package.

Next recall that the $u$-coordinate was constructed as a solution of
the eikonal equation \ref{eq:eikgen2}.
Thus, as seen in the $(u,r,\theta,\phi)$ chart $g^{00}= O_{3}$ which
implies $g_{11} = O_{1}$ and so $a_{11} = 0$.
One may verify this directly by comparing the explicit expression for
$a_{11}$ given in \ref{eq:add} with \ref{eq:mass1}.
We will show later that by making a suitable choice for $y_{2}$ and
$y_{3}$, we can achieve $b_{11}=0$ so that $g_{11} = O_{3}$ as expected.

We now have sufficient notation available to write down the ``outgoing
radiation condition'' of \citet{BVM62} as
\begin{equation}
  \label{eq:bvmorc}
  a_{33}= -a_{22}\sin^{2}\theta, \qquad
  b_{33} = b_{22}\sin^{2}\theta, \qquad
  b_{23} = 0,
\end{equation}
which we shall invoke later.

%%% Local Variables: 
%%% mode: latex
%%% TeX-master: "master"
%%% End: 

\section{The NP tetrad}
\label{sec:tetrad}

Since most recent studies of gravitational radiation use a NP null
tetrad \citep{NP62} we need to introduce one.
The basics of tetrad formalisms are due to \citet{Schou54}.
Many textbooks contain more readable, but often succinct accounts, and
 \citet{Chandra83} chapter 1, section 7, is a good pedagogic
compromise.
With small, but necessary, changes in notation this is summarised in
appendix \ref{sec:appb}.
The specialisation of this approach to the original NP formalism has
been given by  \citet{CW77}.
It turns out that the calculations that we need to do with it become
surprisingly intricate, and so are most conveniently handled using a
computer algebra system.
One of the authors implemented the appendix \ref{sec:appb} formalism,
the other the \citep{CW77} one.
Both of course gave the same results, a useful guard against
programming errors.
Results here are reported for the former.

We use a tetrad of vectors $e_{\alpha}{}^{a}$ and the dual tetrad of
covectors $e^{\alpha}{}_{a}$.  
(The tetrad indices are greek characters and always occur first.)
Tetrad indices are lowered and raised using $\epsilon_{\alpha\beta}$
and $\epsilon^{\alpha\beta}$ where 
\begin{equation*}
  \epsilon_{\alpha\beta} = \epsilon^{\alpha\beta} = \left(
    \begin{array}{cccc}
      0 & 1 & 0 & 0 \\
      1 & 0 & 0 & 0 \\
      0 & 0 & 0 & -1 \\
      0 & 0 & -1 & 0 \\
    \end{array} \right).
\end{equation*}
In NP notation we have 
\begin{equation}
  \label{eq:np1}
  \begin{split}
    e^{0}{}_{a} &= l_{a}, \quad e^{1}{}_{a} = n_{a}, \quad
    e^{2}{}_{a} = m_{a}, \quad e^{3}{}_{a} = \overline{m}_{a}, \\
    e_{0}{}^{a} &= n^{a}, \quad e_{1}{}^{a} = l^{a}, \quad
    e_{2}{}^{a} = -\overline{m}^{a}, \quad e_{3}{}^{a} = -m^{a}. \\
  \end{split}
\end{equation}
We shall require that, to leading order, $l_{a}= u_{,a}$.

Setting $s = 2^{-1/2}$ we write the tetrads as\footnote{In our
  calculations we actually   included one extra term in each of the
  asymptotic expansions below.
  E.g., the first component of $e^{0}{}_{a}$ was written as 
  $$e^{0}{}_{0}=1+c_{00}/r + d_{00}/r^{2} + j_{00}/r^{3} + O_{4}.$$
  These ``junk'' terms show up in our expressions for the connection
  and curvature components.
  In any expression where a junk term occurs we regard all terms of
  that (and any higher order) as being junk, and not computable form
  the data described in section \ref{sec:numerical}.}

\begin{equation}
  \label{eq:npud}
  \begin{split}
    e^{0}{}_{a} &= (1 + c_{00}/r + d_{00}/r^{2} + O_{3}, \quad
    c_{01}/r + d_{01}/r^{2} + O_{3}, \\
    &\qquad c_{02}+ d_{02}/r + O_{2}, \quad c_{03}+ d_{03}/r + O_{2}),\\
    e^{1}{}_{a} &= (\half + c_{10}/r + d_{10}/r^{2} + O_{3}, \quad
    1 + c_{11}/r + d_{11}/r^{2} + O_{3}, \\
    &\qquad c_{12}+ d_{12}/r + O_{2}, \quad c_{13}+ d_{13}/r + O_{2}),\\
    e^{2}{}_{a} &= (c_{20}/r + d_{20}/r^{2} + O_{3}, \quad
    c_{21}/r + d_{21}/r^{2} + O_{3}, \\
    &\qquad -sr + c_{22}+ d_{02}/r + O_{2}, \quad 
    isr\sin\theta + c_{23}+ d_{23}/r + O_{2}),\\
    e^{3}{}_{a} &= (c_{30}/r + d_{30}/r^{2} + O_{3}, \quad
    1 + c_{31}/r + d_{31}/r^{2} + O_{3}, \\
    &\qquad -sr + c_{32}+ d_{32}/r + O_{2}, \quad 
    -isr\sin\theta + c_{33}+ d_{33}/r + O_{2}),\\
  \end{split}
\end{equation}
and 
\begin{equation}
  \label{eq:npdu}
  \begin{split}
    e_{0}^{a} &= (1 + c^{00}/r + d^{00}/r^{2} + O_{3}, \quad
    -\half + c^{01}/r + d^{01}/r^{2} + O_{3}, \\
    &\qquad c^{02}/r^{2} + d^{02}/r^{3} + O_{4}, 
    \quad c^{03}/r^{2} + d^{03}/r^{3} + O_{4}),\\
    e_{1}^{a} &= (c^{10}/r + d^{10}/r^{2} + O_{3}, \quad
    1 + c^{11}/r + d^{11}/r^{2} + O_{3}, \\
    &\qquad c^{12}/r^{2} + d^{12}/r^{3} + O_{4}, 
    \quad c^{13}/r^{2}+ d^{13}/r^{3} + O_{4}),\\
    e_{2}^{a} &= (c^{20}/r + d^{20}/r^{2} + O_{3}, \quad
    c^{21}/r + d^{21}/r^{2} + O_{3}, \\
    &\qquad -s/r + c^{22}/r^{2} + d^{02}/r^{3} + O_{4}, \quad 
    -is\csc\theta/r + c^{23}/r^{2} + d^{23}/r^{3} + O_{4}),\\
    e_{3}^{a} &= (c^{30}/r + d^{30}/r^{2} + O_{3}, \quad
    1 + c^{31}/r + d^{31}/r^{2} + O_{3}, \\
    &\qquad -s/r + c^{32}/r^{2} + d^{32}/r^{3} + O_{4}, \quad 
    is\csc\theta/r + c^{33}/r^{2} + d^{33}/r^{3} + O_{4}).\\
  \end{split}
\end{equation}
Each tetrad contains, at each order, 32 real coefficients.
For although $c_{2n}$ and $c_{3n}$ are complex,
$\overline{c_{2n}}=c_{3n}$ etc.
The relation $e^{\mu}{}_{a}e_{\nu}{}^{a} = \delta^{\mu}{}_{\nu}$
allows one to determine the $c^{mn}$ in terms of the $c_{mn}$ and the
$d^{mn}$ in terms of $c_{mn}$ and $d_{mn}$, reducing the number of
unknowns, at each order, from 32 to 16.
The first set of these is given as equation \ref{eq:cud}  in appendix
\ref{sec:appa} .
The second set is rather lengthy and best generated using a computer
algebra package.

We also have the relation
$\epsilon_{\mu\nu}e^{\mu}{}_{a}e^{\nu}{}_{b}=g_{ab}$, and this enables
us to determine $a_{mn}$ in terms of $c_{mn}$.
Note that there are 10 $a_{mn}$ and 16 real $c_{mn}$.  
Given the tetrad, the metric is uniquely determined.  
But for a given metric  there is a 6-parameter set of tetrads which
give rise to it.
They are of course Lorentz transformations of each other and the
Lorentz group has 6 arbitrary parameters.
We introduce 6 arbitrary first order Lorentz parameters $\alpha_{m}(u,
\theta, \phi)$ and can determine the $c_{mn}$ in terms of the $a_{mn}$
and $\alpha_{m}$. 
There are many different ways of doing this, and one is written down
explicitly as \ref{eq:add1} in appendix \ref{sec:appa}.
We can of course write down the $d_{mn}$ in terms of the $a_{mn}$,
$b_{mn}$, $\alpha_{m}$ and extra second order Lorentz parameters
$\beta_{m}$, but they are rather lengthy and are best generated by
computer algebra. 

%%% Local Variables: 
%%% mode: latex
%%% TeX-master: "master"
%%% End: 

\section{The Curvature Tensors}
\label{sec:curvature}

We now use the tetrads developed in section \ref{sec:tetrad} to
evaluate the Ricci and Weyl curvature tensors using the algorithm
outlined in appendix \ref{sec:appb}.
At each stage we convert all instances of $c^{mn}$ and $d^{mn}$ to
instances of $c_{mn}$ and $d_{mn}$ using \ref{eq:cud} and its second
order analogue.
Then we convert all instances of $c_{mn}$ and $d_{mn}$ to instances of
the metric coefficients $a_{mn}$ and $b_{mn}$ using \ref{eq:add1}
and its second order analogue.
These conversions are implicit and will not be mentioned explicitly again.

We start by looking at the Ricci tensor component 
$R_{11} = R_{ab}e_{1}{}^{a}e_{1}{}^{b}= R_{ab}l^{a}l^{b}$.
We find
\begin{equation}
  \label{eq:ricc11}
  0 = R_{11} = a_{11,u}/r^{2} + O_{3}.
\end{equation}
We chose our chart to ensure that $u$ was approximately a null
coordinate or equivalently $g^{00}=O_{3}$ or equivalently $g_{11}=O_{1}$.  
This means that we have to enforce $a_{11}= 0$, and so the
leading term in $R_{11}$ vanishes. 
We look next at
\begin{equation}
  \label{eq:ricc01}
  0 = R_{01} = R_{ab}e_{0}{}^{a}e_{1}{}^{b}= R_{ab}n^{a}l^{b} = 
  -\half a_{11,uu}/r + O_{2}.
\end{equation}
Again the leading order term vanishes automatically.
Next consider
\begin{equation}
  \label{eq:ricc123} 
  0 = R_{12}+R_{13} = -2sa_{12,u}/r^{2} + O_{3}, \qquad
  0 = R_{12}-R_{13} = -2isa_{13,u}\csc\theta/ r^{2} + O_{3}, \qquad
\end{equation}
where $s = 2^{-1/2}$.
We deduce that
\begin{equation}
  \label{eq:a123}
  a_{12,u} = a_{13,u} = 0.
\end{equation}
Further we can compute 
\begin{equation}
  \label{eq:ricc023} 
  0 = R_{02}+R_{03} = sa_{12,uu}/r + O_{2}, \qquad
  0 = R_{02}-R_{03} = isa_{13,u}\csc\theta/r + O_{2}, \qquad
\end{equation}
If we inspect $R_{23}$ and use \ref{eq:a123}  we find
\begin{equation}
  \label{eq:ricc23}
  0 = R_{23} = \half(a_{22,u} + a_{33,u}\csc^{2}\theta)/r^{2} + O_{3},
\end{equation}
and we deduce that
\begin{equation}
  \label{eq:a22}
  a_{22,u} +  a_{33,u}\csc^{2}\theta = 0.
\end{equation}
Further we may compute
\begin{equation}
  \label{eq:ricc00}
  R_{00} = -\half(a_{22,uu} + a_{33,uu}\csc^{2}\theta )/r + O_{2},
\end{equation}
and we see immediately from \ref{eq:a22} that the leading term vanishes,
and so furnishes no new information.
Finally inspection of the leading $O_{2}$ terms in $R_{22}\pm R_{33}$ reveals
that they vanish automatically because of $a_{11}=0$, \ref{eq:a123}
and \ref{eq:a22}.
Thus we have
\begin{equation}
  \label{eq:ricc223}
  R_{22} + R_{23} = O_{3}, \qquad R_{22} - R_{23} = O_{3}.
\end{equation}
We have found, so far, that the conditions $a_{11}=0$, \ref{eq:a123}
and \ref{eq:a22} imply  that $R_{00}$, $R_{01}$, $R_{02}$ and $R_{03}$
are $O_{2}$ while the other components are $O_{3}$.

At this point we need to examine \ref{eq:a123} more closely.
Using \ref{eq:add} we have
\begin{equation}
  \label{eq:y2z21}
  (h_{02} + h_{12} + y_{2})_{,u} = 0 \qquad
  (h_{03} + h_{13} + z_{2}\sin^{2}\theta)_{,u} = 0. 
\end{equation}
Now we know that the functions $y_{2}$ and $z_{2}$ are arbitrary,
apart form the constraints \ref{eq:yu}, and so \ref{eq:y2z21} implies
\begin{equation}
  \label{eq:y2z22}
  (h_{02} + h_{12})_{,u} = 0 \qquad
  (h_{03} + h_{13})_{,u} = 0. 
\end{equation}
We may therefore choose, consistent with the constraints \ref{eq:yu},
\begin{equation}
  \label{eq:y2z23}
  y_{2} = -(h_{02} + h_{12}), \qquad
  z_{2} = -(h_{03} + h_{13})\csc^{2}\theta,
\end{equation}
which, using \ref{eq:add}  sets 
\begin{equation}
  \label{eq:a1230}
  a_{12} = a_{13} = 0.
\end{equation}

The choice \ref{eq:y2z23} has an added advantage that if we now
express $b_{11}$ in terms of the $h_{ab}$, $k_{ab}$, $m_{1}$ and
$m_{2}$, $y_{2}$ and $z_{2}$ we find that $b_{11}= 0$, 
so that $g_{11}= O_{3}$.
At the same time we can examine $b_{12}$ and $b_{13}$ which are linear
in $y_{3}$ and $z_{3}$ respectively.
By choosing $y_{3}$ and $z_{3}$ appropriately we may arrange
$b_{12}=b_{13}=0$.

This is a convenient point at which to examine the  choice of
a specific Lorentz transformation, in our tetrad, exemplified at
leading order by the parameters $\alpha_{n}$, see e.g., \ref{eq:add1}.
\citet{NP62} chose $u$ to be a null coordinate, $g^{ab}u_{,a}u_{b}=0$
and the covector $l_{a}=u_{,a}$.
For a symmetric connection it follows easily that
$l^{a}{}_{;b}l^{b}=0$.
Even if we set $l_{a}= f(x^{c})u_{,a}$ we find that $l^{a}{}_{;b}l^{b}$
is proportional to $l^{a}$ so that we still have a null geodesic,
albeit not necessarily an affinely parametrised one.
The rest of this paragraph relies on some details of NP formalism
which can be checked swiftly using appendix B of \citet{JMS93}.
Within NP formalism
\begin{equation}
  \label{eq:jms1}
  l^{a}{}_{;b}l^{b}= (\epsilon + \overline \epsilon)l^{a} -
  \bar\kappa m^{a} - \kappa {\overline m}^{a},
 \end{equation}
where $\kappa$ and $\epsilon$ are NP spin coefficients defined below.
Now our coordinate $u$ is only approximately null, and our covector
$l_{a}$ is only approximately its gradient.
Here $\kappa = m^{a}l^{b}l_{a;b}=\gamma_{131}$ (using the
notation of  appendix \ref{sec:appb}) turns out to be $O_{3}$.
However if we choose $\alpha_{4}=\alpha_{5}=\beta_{4}=\beta_{5}=0$,
and anticipate $a_{01}=0$ (see next paragraph), we obtain $\kappa=O_{4}$.
Also $\epsilon + \overline\epsilon = n^{a}l^{b}l_{a;b} = \gamma_{011}$
is $O_{2}$ but if we choose $\alpha_{1}=0$ we find 
$\epsilon + \overline\epsilon$ is $O_{3}$.
We also found $\tau=\gamma_{130} = (\alpha_{2}+i\alpha_{3})/r^{2}+O_{3}$
or $\tau=O_{3}$ if we impose $\alpha_{2}=\alpha_{3}=0$.
At this stage we also choose $\alpha_{6}=\beta_{6}=0$ for reasons 
given below.

Now we need to examine each of the remainder (next order)
terms in \ref{eq:ricc11}, \ref{eq:ricc01}, \ref{eq:ricc123}, 
\ref{eq:ricc023}, \ref{eq:ricc23}, \ref{eq:ricc00} and 
\ref{eq:ricc223}.
For example we now find that the $O_{3}$ terms in $R_{11}$ vanish if
and only if we set $a_{01}=0$, and then $R_{11} = O_{4}$.
This also implies that the $O_{2}$ terms in $R_{01}$ vanish, so that 
$R_{01}=O_{3}$.
We already established that  $R_{12}\pm R_{13} = O_{3}$.  
Setting the leading order terms to zero  furnishes
expressions for $a_{02}$ and $a_{03}$ which we use for subsequent
simplifications. 
Now $R_{12}\pm R_{13} = O_{4}$.  
We find then that our previous estimate \ref{eq:ricc023} 
refines to $R_{02}\pm R_{03} = O_{3}$.
We also need to refine our estimate \ref{eq:ricc23} to $R_{23}=O_{4}$. 
We find $R_{22}\pm R_{23} = O_{3}$, where both $O_{3}$ terms deliver
the same relation relating the $u$-derivatives of 
$a_{22}$, $a_{23}$, $a_{33}$ and $b_{22}$, $b_{23}$, $b_{33}$ which we
save for later use.
In deriving this result we had to choose $\alpha_{6}=\beta_{6}=0$ and
to impose the Bondi outgoing radiation condition \ref{eq:bvmorc}.
Then $R_{22}\pm R_{23} = O_{4}$.
Next we reexamine \ref{eq:ricc00}.
The leading $O_{2}$ term gives us an expression for $a_{00,u}$ which
we store for later use.  Now $R_{00} = O_{3}$.
We have found, so far,  that $R_{00}$, $R_{01}$, $R_{02}$ and $R_{03}$
are $O_{2}$ while the other components are $O_{3}$.

We now try to repeat the procedure of the previous paragraph.
However we find that the $O_{4}$ contribution in $R_{11}$ contains
``junk'' terms, i.e., terms which involve the third order metric
components which we have not been including; see the footnote in
section \ref{sec:tetrad}.
Thus we can obtain no further information from the vacuum field
equation $R_{11}=0$.
Similarly we find that the $O_{3}$ terms in $R_{01}$ contain junk as
do the $O_{4}$ terms in $R_{12}\pm R_{13}$.
The same applies to the $O_{3}$ terms in $R_{02}\pm R_{03}$, the
$O_{4}$ terms in $R_{23}$ and $R_{22}\pm R_{33}$, and finally the 
$O_{3}$ terms in $R_{00}$.
We have therefore exhausted the information available from the vacuum
field equations.

Assuming that we have a vacuum we can switch attention to the Weyl
tensor, and we compute first
\begin{equation}
  \label{eq:psi4a}
  \Psi_{4} = 
  R_{0202} = \left[
    \quarter(a_{22,uu} - a_{33,uu}\csc^{2}\theta) + 
    \half i a_{23,uu}\csc\theta \right]/r + O_{2}.
\end{equation}
Using \ref{eq:a22} we may rewrite this as
\begin{equation}
  \label{eq:psi4b}
  \Psi_{4} = \mathcal{N}_{,u}/r + O_{2},
\end{equation}
where 
\begin{equation}
  \label{eq:bnews}
  \mathcal{N} = \half(a_{22} + ia_{23}\csc\theta)_{,u}
\end{equation}
is the \emph{Bondi news function} \citep{BVM62}.

This is a highly satisfactory result which, in spite of our rather ad
hoc chart and tetrad, mimics the treatment of \citet{BVM62} and
\citet{NU62}.
Further we see that it is linear in the $a_{mn}$ and so should appear
 in linearized theory.
Also it does not involve the Lorentz parameters $\alpha_{n}$ and
$\beta_{n}$ and so is tetrad-invariant (for tetrads which are
asymptotically Minkowskian).
The remainder term in in \ref{eq:psi4b} contains some $O_{2}$ terms and
junk $O_{3}$ terms.

Next consider
\begin{equation}
  \label{eq:psi3a}
  \Psi_{3} = R_{0120} = \Psi_{3}^{(2)}/r^{2} + O_{3},
\end{equation}
where 
\begin{equation}
  \label{eq:psi3b}
  \Psi_{3}^{(2)} = 
  2^{-1/2}(\mathcal{N}_{,\theta}-i\mathcal{N}_{,\phi}\csc\theta +  
  \mathcal{N}\cot\theta) + 
(\alpha_{4} + i\alpha_{5})\mathcal{N}_{,u}.
\end{equation}
Note first that the $r$-dependence is precisely what one would have
expected from the peeling property.
The first  term in  the coefficient $\Psi_{3}^{(2)}$ is linear
and would have been predicted within linearized theory.
However the second term is nonlinear for  it depends on the
$\alpha_{n}$ which determine the infinitesimal Lorentz transformation
of the NP tetrads \ref{eq:npud} and \ref{eq:npdu}.
This is to be expected.
The NP tetrad used by \citet{NU62} was chosen very specifically, while
here we are considering a class of tetrads
infinitesimally close to the Minkowski one.
If we were to restrict  attention to the subclass of tetrads where
$\alpha_{4}=\alpha_{5}=0$ then our result would be consistent with
linearized theory.
On the other hand another choice of $a_{4}+i\alpha_{5}$ would give
$\Psi_{3}^{(2)} = 0$.
The remainder term in \ref{eq:psi3a} is junk.

Next we find that 
\begin{equation}
  \label{eq:psi2a}
  \Psi_{2} = R_{1320} = \Psi_{2}^{(3)}/r^{3} + O_{4},
\end{equation}
where the remainder term is junk.
We will return to the leading term shortly.

We find next that 
\begin{equation}
  \label{eq:psi1}
  \Psi_{1} = R_{0113} = \Psi_{1}^{(4)}/r^{4} + O_{5},
\end{equation}
The coefficient $\Psi_{1}^{(4)}$ contains nonlinear terms, but we are
unable to determine it precisely because it also contains junk terms. 
The peeling property is still holding though.

The peeling property would demand that $\Psi_{0} = R_{1313}$ should be
$O_{5}$. 
However we find
\begin{equation}
  \label{eq:psi0a}
  \Psi_{0} = \Psi_{0}^{(4)}/r^{4} + O_{5},
\end{equation}
where
\begin{equation}
  \label{eq:psi0b}
  \Psi_{0}^{(4)} = \tfrac{1}{8}[
  (a_{22}+a_{33}\csc^{2}\theta)(a_{22}-2ia_{23}\csc\theta  - 
  a_{33}\csc^{2}\theta) + 4(b_{22}-b_{33}\csc^{2}\theta - 2ib_{23}\csc\theta)].
\end{equation}
But we have not made the restrictions that were imposed by
\citet{BVM62} or \citet{NU62} to ensure peeling.
The latter restriction was to \emph{demand} $\Psi_{0}^{(4)} =0$.
The former restriction was that the ``outgoing radiation condition'',
\citep{BVM62}, held.
In our notation this condition is , see \ref{eq:bvmorc},
\begin{equation}
  \label{eq:orc1}
  a_{22} = -a_{33}\csc^{2}\theta, \qquad
  b_{22} = b_{33}\csc^{2}\theta, \qquad  b_{23} = 0.
\end{equation}
Examining \ref{eq:psi0b} we see that imposing the outgoing radiation
condition \ref{eq:orc1} ensures $\Psi_{0}^{(4)} =0$, a result first
obtained by \citet{Kroon98}.
With one or other condition we have
\begin{equation}
  \label{eq:psi0c}
  \Psi_{0} = \Psi_{0}^{(5)}/r^{5} + O_{6},
\end{equation}
but the coefficient $\Psi_{0}^{(5)}$ contains junk terms and so we
cannot evaluate it.
(It also contains nonlinear terms not predicted by linearized theory.)

To summarize: if we impose the outgoing radiation condition
\ref{eq:orc1} then we obtain the peeling property, and we can obtain
explicitly the leading terms in $\Psi_{4}$, $\Psi_{3}$, $\Psi_{2}$, but
not those for $\Psi_{1}$ and $\Psi_{0}$ because they contain junk
terms.

We now return to the discussion of $\Psi_{2}$ given by \ref{eq:psi2a}.
The leading term coefficient is 
\begin{equation}
  \label{eq:psi2b}
  \begin{split}
    \Psi_{2}^{(3)} =& \half a_{00} - \quarter ia_{23}\csc^{3}\theta +
    \half b_{22,u} +
    \quarter(a_{22}+ia_{23}\csc\theta )
    (a_{22,u} - ia_{23,u}\csc\theta ) +\\
    &\quad \quarter i (a_{23,\theta} - 2a_{22,\phi})\cot\theta\csc\theta -
    \half ia_{22,\theta\phi}\csc\theta  +
    \quarter i (a_{23,\theta\theta} - 
    \csc^{2}\theta a_{23,\phi\phi})\csc\theta.
  \end{split}
\end{equation}
Here we have fixed the Lorentz parameters, as described earlier, and
are imposing the outgoing radiation condition \ref{eq:bvmorc}.

Now the \emph{Bondi mass} $M_{B}(u)$ can be  defined by, \citep{BVM62},
\citep{NU62}, \citep{JMS89},
\begin{equation}
  \label{eq:bmassdef}
  4\pi M_{B}(u) = -\lim_{r\rightarrow\infty}\int_{S(u,r)}
  r^{3}(\Psi_{2} + \sigma\lambda)\,\sin\theta\,d\theta\,d\phi,
\end{equation}
where the integral is over the 2-surface $S(u,r)$ given by $u=const.$
and $r=const.$
Here $\sigma$ and $\lambda$ are NP spin coefficients given by
\begin{equation}
  \label{eq:siglam1}
  \sigma = m^{a}\delta l_{a} = \gamma_{313} = \sigma^{(2)}/r^{2} +
  O_{3}, \qquad
  \lambda = n^{a}{\bar \delta}{\overline m}_{a} = 
  \gamma_{022} = \lambda^{(1)}/r + O_{2},
\end{equation}
where
\begin{equation}
  \label{eq:siglam2}
  \sigma^{(2)} =  -\half(a_{22}-ia_{23}\csc\theta ), \qquad
  \lambda^{(1)} = -\half (a_{22,u}+ia_{23,u}\csc\theta ).
\end{equation}
Taking the limit in \ref{eq:bmassdef} we have
\begin{equation}
  \label{eq:bmassdef1}
  4\pi M_{B}(u) = -\int_{S(u,1)}(\Psi_{2}^{(3)} + 
  \sigma^{(2)}\lambda^{(1)})\,\sin\theta\,d\theta\,d\phi.
\end{equation}

Of course the formula \ref{eq:bmassdef} is only valid in a specially
chosen \emph{Bondi frame}.
The generalization to an arbitrary NP frame is discussed in
\citet{JMS89}.
In the large $r$ limit our frame differs from the Bondi one by a
Lorentz transformation which is close to the identity.
A 2-parameter subgroup of the Lorentz group consists of ``boosts'' and
``spins'' 
\begin{equation}
\label{eq:boostspin}
l \rightarrow a^{2}l,\quad
n \rightarrow a^{-2}n,\quad
m \rightarrow e^{i\psi}m,  
\end{equation}
where $a$ and $\psi$ are real.
Using the formulae in appendix B of \cite{JMS93} it is easy to verify
that the integrand of \ref{eq:bmassdef1} is invariant under boosts and
spins.
Next consider a 2-parameter subgroup of ``null rotations about $l$''
given by
\begin{equation}
  \label{eq:lrot1}
  l \rightarrow l,\quad, m \rightarrow m + {\bar c}l,\quad
  n \rightarrow n + cm + {\bar c}{\overline m} + c{\bar c}l,
\end{equation}
where $c$ is complex.
Under such a transformation
\begin{equation}
  \label{eq:lrot2}
  \begin{split}
    &\Psi_{2}\rightarrow \Psi_{2} + 2c\Psi_{1} + c^{2}\Psi_{0}, \qquad
    \sigma \rightarrow \sigma + {\bar c}\kappa,\\
    &\lambda\rightarrow\lambda + c\pi + 2c\alpha +
    c^{2}(\rho+2\epsilon) + c^{3}\kappa + cl^{a}c_{,a} + 
    {\overline m}^{a}c_{,a}.
  \end{split}
\end{equation}
We expect $c = O_{1}$ and the NP scalars $\alpha$, $\pi$, $\rho$ and
$\epsilon$ are all $O_{1}$.
Thus the integrand of \ref{eq:bmassdef1} is not changed.
We should also consider null rotations about $n$ given by
\begin{equation}
  \label{eq:nrot1}
  n \rightarrow n,\quad, m \rightarrow m + {\bar c}n,\quad
  l \rightarrow l + cm + {\bar c}{\overline m} + c{\bar c}n,
\end{equation}
so that
\begin{equation}
  \label{eq:nrot2}
  \begin{split}
    &\Psi_{2}\rightarrow \Psi_{2} + 2c\Psi_{3} + c^{2}\Psi_{4}, \qquad
    \lambda \rightarrow \lambda + {\bar c}\nu,\\
    &\sigma\rightarrow\sigma + c\tau + 2c\beta +
    c^{2}(\mu+2\gamma) + c^{3}\nu + cn^{a}c_{,a} +  m^{a}c_{,a}.
  \end{split}
\end{equation}
Now we have taken great care to ensure that $l$ is almost geodesic (
$\kappa=O_{4}$) and almost affinely parametrised 
($\epsilon + \bar\epsilon=O_{3}$) and so we should only consider the
transformation \ref{eq:nrot1} where $c=O_{3}$.
Under this restriction the integrand of \ref{eq:bmassdef1} is not
changed.
Thus the formula \ref{eq:bmassdef1} evaluated in our frame does indeed
give the Bondi mass to leading order.
Next note that
\begin{equation}
  \label{eq:bmassim}
  \begin{split}
  Im(\Psi_{2}^{(3)} + \sigma^{(2)}\lambda^{(1)}) =&
  -\quarter a_{23}\csc^{3}\theta + 
  \quarter a_{23,\theta}\csc\theta\cot\theta +
  \quarter a_{23,\theta\theta}\csc\theta  -\\
  & \qquad \half  a_{22,\phi}\csc\theta\cot\theta -
  \half a_{22,\theta\phi}\csc\theta  -
  \quarter a_{23,\phi\phi}\csc^{3}\theta.
  \end{split}
\end{equation}
When we integrate this over the unit sphere the terms in the second
line give zero since their contribution to the integrand is
$2\pi$-periodic in $\phi$. 
Those in the first line contribute
\begin{equation*}
  \half\pi\left[\csc\theta(\sin\theta a_{23})_{,\theta}\right]^{\pi}_{0}.
\end{equation*}
Now $a_{23}$ must scale like $\sin^{2}\theta$ at the end points or
else the integrand is singular.  It follows that the Bondi mass must
be real, and
\begin{equation}
  \label{eq:bmassre}
  4\pi M_{B}(u) = -\half\int_{0}^{\pi}(
  a_{00} + b_{22,u} + a_{22}a_{22,u} +  a_{23}a_{23,u}\csc^{2}\theta)
  \sin\theta\,d\theta\,d\phi.
\end{equation}

Finally there is a standard result, \citep{BVM62}, \citep{NU62},
\citep{JMS89}, for the rate of decrease of the Bondi mass
\begin{equation}
  \label{eq:bmassloss}
  \begin{split}
    4\pi\frac{dM_{B}}{du}(u) &= -\int_{S(u,1)} |\mathcal{N}|^{2}
    \sin\theta\,d\theta\,d\phi\\
    &= -\quarter \int_{S(u,1)}\left((a_{22,u})^{2} + 
      (a_{23,u})^{2}\csc^{2}\theta\right)\sin\theta\,d\theta\,d\phi,
  \end{split}
\end{equation}
demonstrating the well-known result $dM_{B}/du\leqslant 0$, the
Bondi mass decreases as energy is radiated away, a result not
deducible in linearized theory.
Although \ref{eq:bmassloss} was originally derived in a special Bondi
frame, it too holds in our approximate Bondi one, at least to leading
order.
We should empasize that although the outgoing radiation condition was
used in the derivation of \ref{eq:bmassre}, the mass loss formula
\ref{eq:bmassloss} holds without the need for this restriction.
%%% Local Variables: 
%%% mode: latex
%%% TeX-master: "master"
%%% End: 

\section{Implications for numerical relativity}
\label{sec:numrec}

At first glance the formalism we set up to carry out this study may
seem to be cumbersome, but it has the advantage the the results can be
translated back into the $X^{a}=(T,R,\Theta,\Phi)$ chart, after which
the theoretical chart $x^{a}=(u,r,\theta,\phi)$ can be discarded.

We chose the $x^{a}$ chart so that the metric coefficients $a_{11}$
and $b_{11}$ vanished as well as $a_{12}$ and $a_{13}$.
Now
\begin{equation}
  \label{eq:partial}
  \begin{split}
    \left(\frac{\partial\;\;}{\partial u}\right)_{r} &=
    \left(\frac{\partial T}{\partial u}\right)_{r}
    \left(\frac{\partial\;\;}{\partial T}\right)_{R} + 
    \left(\frac{\partial R}{\partial u}\right)_{r}
    \left(\frac{\partial\;\;}{\partial R}\right)_{T} \\
    &= (1-q_{1}r^{-1})\left(\frac{\partial\;\;}{\partial T}\right)_{R}+ O_{2},\\
  \end{split}
\end{equation}
using \ref{eq:jac1}.
In principle the function $q_{1}$ is arbitrary.
But the vacuum field equations implied $a_{01}=0$, and then equations
\ref{eq:add} imply
\begin{equation}
  \label{eq:q1}
  q_{1} = \half(h_{00}-h_{11}).
\end{equation}
The vacuum field equations imply \ref{eq:a22}
\begin{equation*}
  a_{22,u} + a_{33,u}\csc^{2}\theta.
\end{equation*}
Using \ref{eq:add} and \ref{eq:y2z23} we have, to leading order
\begin{equation}
  \label{eq:eg}
  [h_{22}+2h_{02,\Theta}+2h_{12,\Theta}]_{,T} + 
  [h_{33}+(2h_{03,\Phi}+2h_{13,\Phi})\csc^{2}\Theta]_{,T} \csc^{2}\Theta = 0.
\end{equation}
Other vacuum conditions can be handled in a similar way.

In order to discuss the Bondi news function and Bondi mass, it is
convenient to introduce some auxiliary functions in the numerical chart,
\begin{equation}
  \label{eq:aux}
  \begin{split}
    \mathcal{W} =& h_{03} + h_{13},\\
    \mathcal{A} =& h_{22}+ 2(h_{02,\Theta}+h_{12,\Theta}),\\
    \mathcal{B} =& h_{23} + h_{02,\Phi} + h_{12,\Phi} +
    \mathcal{W}_{,\Theta} - 2\mathcal{W}\cot\Theta, \\
    \mathcal{C} =& k_{22} + 2\mathcal{W}h_{23}\cot\Theta\csc^{2}\Theta
    - 4\mathcal{W}^{2}\cot\Theta\csc^{2}\Theta +
    (k_{02}+k_{12})_{,\Theta} -\\
    &\quad (\half h_{11}+h_{22})h_{02,\Theta} -
    ((\half h_{00}+h_{01})h_{02})_{,\Theta} +
    (4\mathcal{W}\cot\Theta -
    h_{23})\mathcal{W}_{,\Theta}\csc^{2}\Theta - \\
    &\quad \half h_{02}h_{11,\Theta} - h_{22}h_{12,\Theta} + 
    (h_{02}+h_{12})h_{22,\Theta} +
    \mathcal{W}h_{23,\Theta}\csc^{2}\Theta -
    (\mathcal{W}_{,\Theta})^{2}\csc^{2}\Theta
  \end{split}
\end{equation}
which should be readily available according to the assumptions in
section \ref{sec:numerical}.

Then the leading term in the Bondi news function given by
\ref{eq:bnews} becomes 
\begin{equation}
  \label{eq:bnewsnum}
  \mathcal{N} = \mathcal{A}_{,T} + i\mathcal{B}_{,T}\csc\Theta,
\end{equation}
whose calculation  might require some sophistication, although there
is no reference to the intermediary $x^{a}=(u,r,\theta,\phi)$ chart.
Because the news function is linear in the $h_{ab}$ and their
derivatives, it could have been calculated within linearized theory.

In particular the formula \ref{eq:bmassre} for the Bondi mass
$M_{B}(u)$ translates into
\begin{equation}
  \label{eq:bmassnum}
  \begin{split}
    M_{B}(T-R) =&
    -\frac{1}{8\pi}\int_{\Phi=0}^{2\pi}\int_{\Theta=0}^{\pi}
     \Big(h_{11} + \mathcal{A}\mathcal{A}_{,T}
     + \mathcal{B}\mathcal{B}_{,T}\csc^{2}\Theta + \mathcal{C}_{,T} \Big)
       \,\sin\Theta\,d\Theta d\Phi.
   \end{split}
\end{equation}

In a similar way the formula \ref{eq:bmassloss} for the rate of change
of $M_{B}$ at fixed large $R$ is
\begin{equation}
  \label{eq:blossnum}
  \begin{split}
    {\dot M}_{B}(T-R) = &
    -\frac{1}{16\pi}\int_{\Phi=0}^{2\pi}\int_{\Theta=0}^{\pi}\Big(
    (\mathcal{A}_{,T})^{2} + (\mathcal{B}_{,T})^{2}\csc^{2}\Theta
    \Big)\,\sin\Theta\,d\Theta d\Phi.
   \end{split}
\end{equation}

Working in linearized theory numerical relativists often take the
leading terms in $\cal{A}$ and $\cal{B}$ to represent the
``gravitational waveforms'' $h_+$ and $h_{\times}$. Equations \ref{eq:aux}
suggest that other terms are present, even in linearized theory.

Again we emphasize that the intermediate $x^{a}$ chart does not
intrude---the formulae \ref{eq:bnewsnum}, \ref{eq:bmassnum} and
\ref{eq:blossnum} apply in the numerical $X^{a}=(T,R,\Theta,\Phi)$
chart.
Only the first can be deduced from linearized theory.

Why are these formulae so complicated, when compared with the original
papers, \citet{BVM62} and \citet{NU62}, or even the formulae in
section \ref{sec:curvature}?
Well the coordinates and tetrads of the originals were very carefully
chosen to simplify the problem, and much of this paper has been spent
building the relationship between the numerical relativist's 
$X^{a}=(T,R,\Theta, \Phi)$ chart and the $x^{a}=(u,r,\theta,\phi)$
chart and adapted tetrad used in this paper, in which the formulae look
almost as simple as in the original approaches.
One way to avoid the complexity is to design a numerical approach
based on Penrose's geometrical approach \citep{RP63}, but that brings
in different problems and complexities.
%%% Local Variables: 
%%% mode: latex
%%% TeX-master: "master"
%%% End: 

\appendix
\section{Computational Details}
\label{sec:appa}
\renewcommand{\theequation}{A\arabic{equation}}
\setcounter{equation}{0}

The $h^{ab}$ and $k^{ab}$ occurring in \ref{eq:gtruu} are given by
\begin{gather}
  h^{00} = -h_{00}, \quad h^{01} = h_{01}, \quad h^{02} = h_{02}, 
  \quad h^{03} = \csc^{2}\theta h_{03}, \quad h^{11} = -h_{11},
  \quad h^{12} = -h_{12}, \notag \\
   h^{13} = -\csc^{2}\theta h_{13}, 
  \quad h^{22} = -h_{22}, \quad h^{23} = -\csc^{2}\theta h_{23}, 
  \quad h^{33} = -\csc^{4}\theta h_{33}, \label{eq:huu}
\end{gather}
and
\begin{equation}
  \label{eq:kuu}
  \begin{split}
    k^{00} &= -k_{00} + h_{00}^2 -  h_{01}^2 - h_{02}^2 - 
    \csc^{2}\theta h_{03}^2,\\
    k^{01} &= k_{01} - h_{00}h_{01} +  h_{01}h_{11} +
    h_{02}h_{12} + \csc^{2}\theta h_{03}h_{23}, \\
    k^{02} &= k_{02} -h_{00}h_{02}+h_{01}h_{12} + h_{02}h_{22} +
    \csc^{2}\theta h_{03}h_{23}, \\
    k^{03} &= \csc^{2}\theta(k_{03} - h_{00}h_{03} + h_{01}h_{13} +
    h_{02}h_{23} + \csc^{2}\theta h_{03}h_{33}), \\
    k^{11} &= -k_{11} + h_{01}^2 - h_{11}^2 - h_{12}^2 -
    \csc^{2}\theta h_{13}^2, \\
    k^{12} &= -k_{12} + h_{01}h_{02} - h_{11}h_{12} - h_{12}h_{22} -
    \csc^{2}\theta h_{13}h_{23}, \\
    k^{13} &= \csc^{2}\theta(-k_{13} + h_{01}h_{03} - h_{11}h_{13} - 
    h_{12}h_{23} - \csc^{2}\theta h_{13}h_{33}),\\
    k^{22} &= -k_{22} +h_{02}^2 - h_{12}^2 - h_{22}^2 -
   \csc^{2}\theta h_{23}^2, \\
    k^{23} &= \csc^{2}\theta(-k_{23} + h_{02}h_{03} - h_{12}h_{13} -
    h_{22}h_{23} - \csc^{2}\theta h_{23}h_{33}), \\
    k^{33} &= \csc^{4}\theta(-k_{33} + h_{03}^2 - h_{13}^2 -
    h_{23}^2 - \csc^{2}\theta h_{33}^2).
\end{split}
\end{equation}

The $a_{ab}$ occurring in \ref{eq:gurdd} (after imposing the
conditions \ref{eq:yu}) are given by
\begin{equation}
  \label{eq:add}
  \begin{split}
    a_{00} &= h_{00} - 2q_{1},\\
    a_{01} &= h_{00} + h_{01} + 2m_{1} -q_{1},\\
    a_{02} &= h_{02} - {y_{3}}_{,u}, \\
    a_{03} &= h_{03} - {z_{3}}_{,u}\sin^{2}\theta ,\\
    a_{11} &= h_{00} + 2h_{01} + h_{11} + 4m_{1}, \\
    a_{12} &= h_{02} + h_{12} + y_{2},\\
    a_{13} &= h_{03} + h_{13} + z_{2}\sin^{2}\theta ,\\ 
    a_{22} &= h_{22} - 2{y_{2}}_{,\theta},\\
    a_{23} &= h_{23} - {y_{2}}_{,\phi} - {z_{2}}_{,\theta}\sin^{2}\theta ,\\
    a_{33} &= h_{33} - 2{z_{2}}_{,\phi}.
  \end{split}
\end{equation}

The relation between the tetrad components $c^{mn}$ and the $c_{mn}$
is (here $s=2^{-1/2}$)
\begin{equation}
  \label{eq:cud}
  \begin{split}
    c^{00} &= \half c_{01} - c_{00},\\
    c^{01} &= \half c_{00} - \quarter c_{01} - c_{10} + \half c_{11},\\
    c^{02} &= s(c_{20} + c_{30}) - \half s(c_{21} + c_{31}),\\
    c^{03} &= is[(c_{20} - c_{30}) - \half s(c_{21} - c_{31})\csc\theta],\\
    c^{10} &= -c_{01},\\
    c^{11} &= \half c_{01} - c_{11},\\
    c^{12} &= s(c_{21} + c_{31}),\\
    c^{13} &= is(c_{21} - c_{31})\csc\theta,\\
    c^{20} &= s(c_{02} + ic_{03}\csc\theta),\\
    c^{21} &= s(c_{12} - \half c_{02}) - is(\half c_{03} - c_{13})\csc\theta,\\
    c^{22} &= -\half(c_{22}+c_{32}) - \half i(c_{23}+c_{33})\csc\theta,\\
    c^{23} &= -\half i(c_{22}-c_{32})\csc\theta +
    \half(c_{23}-c_{33})\csc^{2}\theta, \\
    c^{30} &= s(c_{02} - ic_{03}\csc\theta),\\
    c^{31} &= s(c_{12} - \half c_{02}) +
    is(\half c_{03} - c_{13})\csc\theta,\\
    c^{32} &= -\half(c_{22}+c_{32}) + \half i(c_{23}+c_{33})\csc\theta,\\
    c^{33} &= -\half i(c_{22}-c_{32})\csc\theta -
    \half(c_{23}-c_{33})\csc^{2}\theta. \\
  \end{split}
\end{equation}

One possible relation between the tetrad coefficients $c_{mn}$ and the
metric coefficients $a_{mn}$ and Lorentz parameters $\alpha_{m}$ is
\begin{equation}
  \label{eq:add1}
  \begin{split}
    c_{00} &= \alpha_{1},\\
    c_{01} &= \half a_{11}, \\
    c_{02} &= a_{12} - 2s\alpha_{4},\\
    c_{03} &= a_{13} + 2s\alpha_{5}\sin\theta ,\\
    c_{10} &= \half a_{00} - \half \alpha_{1},\\
    c_{11} &= a_{01} - \quarter a_{11} - \alpha_{1},\\
    c_{12} &= a_{02} - \half a_{12} - s(2\alpha_{2} - \alpha_{4}),\\
    c_{13} &= a_{03} - \half a_{13} + s(2\alpha_{3}-\alpha_{5})\sin\theta,\\
    c_{20} &= \alpha_{2} +i\alpha_{3},\\
    c_{21} &= \alpha_{4} +i\alpha_{5},\\
    c_{22} &= \half s a_{22} - is(a_{23}-\alpha_{6})\csc\theta,\\
    c_{23} &= s\alpha_{6} - \half isa_{33}\csc\theta,\\
    c_{30} &= \alpha_{2} - i\alpha_{3},\\
    c_{31} &= \alpha_{4} - i\alpha_{5},\\
    c_{32} &= \half s a_{22} + is(a_{23}-\alpha_{6})\csc\theta,\\
    c_{33} &= s\alpha_{6} + \half isa_{33}\csc\theta,\\
  \end{split}
\end{equation}
where $s=2^{-1/2}$.
Other representations are possible.
%%% Local Variables: 
%%% mode: latex
%%% TeX-master: t
%%% End: 

\section{Tetrad formalism}
\label{sec:appb}
\renewcommand{\theequation}{B\arabic{equation}}
\setcounter{equation}{0}

Here we define the notation and summarize the results.
The reader to whom this material is unfamiliar should consult
introductory material e.g., \citet{Chandra83} chapter 1, section 7.

In this section $a, b, c,\ldots$ are coordinate indices while 
$\alpha, \beta, \gamma, \ldots$ are tetrad indices.

At each space-time point $P$ we introduce a basis of vectors
\begin{equation}
  \label{eq:bas1}
  e_{\alpha}{}^{a}, \qquad \alpha\in[0,3], \quad a\in [0,3].
\end{equation}
Then the matrix
\begin{equation*}
  e_{\alpha}{}^{a} = \left (
    \begin{array}{ccc}
      e_{0}{}^{0} & e_{0}{}^{1} & \ldots\\
      e_{1}{}^{0} & e_{1}{}^{1} & \ldots\\
      \vdots & \vdots & \ddots 
    \end{array} \right)
\end{equation*}
is non-singular and we denote its inverse by $e^{\alpha}{}_{a}$.
Thus
\begin{equation}
  \label{eq:inv1}
  e_{\alpha}{}^{a}e^{\beta}{}_{a} = \delta_{\alpha}{}^{\beta}, \qquad
  e_{\alpha}{}^{a}e^{\alpha}{}_{b} = \delta^{a}{}_{b}.
\end{equation}
The $e^{\alpha}{}_{a}$ represent the dual basis of covectors.
As usual chart indices are lowered (raised) using $g_{ab}$($g^{ab}$).

An additional assumption made here is that
\begin{equation}
  \label{eq:eps1}
  \epsilon_{\alpha\beta} = g_{ab}\,e_{\alpha}{}^{a}e_{\beta}{}^{b}
\end{equation}
is a \textbf{constant} symmetric matrix with inverse
$\epsilon^{\alpha\beta}$.
Thus
\begin{equation}
  \label{eq:eps2}
  \epsilon_{\alpha\beta}\,\epsilon^{\beta\gamma} = \delta_{\alpha}{}^{\gamma}.
\end{equation}
The choice $\epsilon_{\alpha\beta}=\text{diag}(1,-1,-1,-1)$ gives an
\emph{orthonormal tetrad}, but here we choose 
\begin{equation}
  \label{eq:npt1}
  \epsilon_{\alpha\beta} = \epsilon^{\alpha\beta} = \left(
    \begin{array}{cccc}
      0 & 1 & 0 & 0 \\
      1 & 0 & 0 & 0 \\
      0 & 0 & 0 & -1 \\
      0 & 0 & -1 & 0 \\
    \end{array} \right),
\end{equation}
which gives a NP tetrad, \cite{NP62}.

Then it is easy to see that 
\begin{equation}
  \label{eq:inv2}
  \epsilon_{\alpha\beta}\,e^{\beta}{}_{a} = e_{\alpha a}, \qquad
  \epsilon^{\alpha\beta}\,e_{\beta}{}^{a} = e^{\alpha a},
\end{equation}
so that tetrad indices are lowered (raised) using 
$\epsilon_{\alpha\beta}$ ($\epsilon^{\alpha\beta}$).

The \emph{Ricci rotation coefficients} $\gamma_{\lambda\mu\nu}$ are
defined via
\begin{equation}
  \label{eq:ricci1}
  e_{\mu b;c} = \gamma_{\lambda\mu\nu} e^{\lambda}{}_{b}e^{\nu}{}_{c},
\end{equation}
where the metric covariant derivative has been used,
and since $\epsilon_{\alpha\beta}$ is constant we must have
$\gamma_{\lambda\mu\nu} = \gamma_{[\lambda\mu]\nu}$.

The tetrad \emph{structure constants} $C^{\gamma}{}_{\alpha\beta}$ are
defined via
\begin{equation}
  \label{eq:struct1}
  [e_{\alpha}, e_{\beta}] = C^{\gamma}{}_{\alpha\beta}e_{\gamma},
\end{equation}
and clearly $C^{\gamma}{}_{\alpha\beta} = C^{\gamma}{}_{[\alpha\beta]}$.
If we let \ref{eq:struct1} act on a scalar function $f$, note that the
metric connection is symmetric, and use \ref{eq:ricci1}, then it is
easy to see that
\begin{equation}
  \label{eq:struct2}
  C^{\gamma}{}_{\alpha\beta} = \gamma^{\gamma}{}_{\beta\alpha} - 
  \gamma^{\gamma}{}_{\alpha\beta},
\end{equation}
which implies
\begin{equation}
  \label{eq:struct3}
  \gamma_{\lambda\mu\nu} = \half(C_{\nu\lambda\mu} - C_{\lambda\mu\nu}
  -   C_{\mu\nu\lambda}).
\end{equation}

It is important to realise that the $C^{\lambda}{}_{\mu\nu}$ do not
involve the connection.
For \ref{eq:struct2}, \ref{eq:ricci1} and the fact that the connection
is symmetric means that
\begin{equation}
  \label{eq:struct4}
  C^{\lambda}{}_{\mu\nu} = e^{\lambda}{}_{a,b}
  (e_{\mu}{}^{a}e_{\nu}{}^{b} - e_{\mu}{}^{b}e_{\nu}{}^{a}).
\end{equation}

Next the Ricci identity applied to $e^{\alpha}{}_{b}$ gives
\begin{equation}
  \label{eq:riemt}
  R_{\alpha\beta\gamma\delta} = \gamma_{\alpha\beta\gamma,a}e_{\delta}{}^{a}  -
  \gamma_{\alpha\beta\delta,a}e_{\gamma}{}^{a} +
  \gamma_{\alpha\beta\epsilon}C^{\epsilon}{}_{\gamma\delta} +
  \gamma^{\epsilon}{}_{\alpha\gamma}\gamma_{\epsilon\beta\delta} - 
  \gamma^{\epsilon}{}_{\alpha\delta}\gamma_{\epsilon\beta\eta}, 
\end{equation}
and finally
\begin{equation}
  \label{eq:riccit}
  R_{\alpha\gamma} = \epsilon^{\beta\delta}R_{\alpha\beta\gamma\delta}.
\end{equation}
Note that throughout the paper we use these forms for the curvature
tensors.
Thus $R_{12}$ means $R_{\alpha\beta}$ with $\alpha=1$ and $\beta=2$,
which is not the same as $R_{ab}$ with $a=1$ and $b=2$.

Our algorithm starts from the sets $\{e_{\alpha}{}^{a}\}$ and 
$\{e^{\alpha}{}_{a}\}$.
We compute $C^{\alpha}{}_{\beta\gamma}$ from \ref{eq:struct4} and of
course $C_{\gamma\alpha\beta} = \epsilon_{\gamma\delta}C^{\delta}_{\alpha\beta}$.
Next we compute $\gamma_{\alpha\beta\gamma}$ from \ref{eq:struct3}
and finally the curvature tensors from \ref{eq:riemt} and \ref{eq:riccit}.
Although this looks ponderous it can easily be automated using a
computer algebra system.

%%% Local Variables: 
%%% mode: latex
%%% TeX-master: "master"
%%% End: 

%%% Local Variables: 
%%% mode: latex
%%% TeX-master: "master"
%%% End: 

\end{document}